# Gate-tunable weak localization and magnetoconductance in La-doped $SrSnO_3$ thin-film devices

Long Cheng[1,2], Junghyun Koo[1], Dongho Won[3], Donghwan Kim[4], Chengyu Zhu[1], Bharat Jalan[4], Kiyoung Lee[3], Weideng Sun[1,a)], and Gang Qiu[1,a)]

[1]Department of Electrical and Computer Engineering, University of Minnesota, Twin Cities, Minneapolis, MN, 55455, USA

[2]Department of Physics and Astronomy, Carleton College, Northfield, MN, 55057, USA

[3]Department of Materials Science & Engineering, Hongik University, Seoul, Republic of Korea

[4]Department of Chemical Engineering and Materials Science, University of Minnesota, Twin Cities, Minneapolis, MN, 55455, USA

[a)]Corresponding author. E-mail: gqiu@umn.edu, weidengs@umn.edu

## Abstract

Understanding electrical control of disorder-dominated transport in ultrawide bandgap oxide semiconductors is important for developing gate-modulated electronic devices. Here, we investigate temperature-, angle-, and gate-dependent magnetotransport in top-gated La-doped $SrSnO_3$ thin films. $Al_2O_3$ dielectric capping layer enhances carrier density and mobility, consistent with dielectric-induced modulation doping and improved screening of charged impurities. At low temperatures, the capped channel exhibits large positive magnetoconductance characteristic of weak localization (WL), arising from the magnetic-field suppression of constructive interference between time-reversed electron trajectories. The extracted phase-coherence length increases from 23.3 nm at 20 K to 112.1 nm at 1.7 K. Angle-dependent measurements reveal that the low-field magnetoconductance is dominated primarily by the perpendicular component of the magnetic field, supporting predominantly in-plane phase-coherent transport. The WL feature can further be tuned by a top gate. By systematically analyzing the gate dependence of the phase-coherence length, the electron-electron interaction is the dominant phase-relaxation mechanism at 1.7 K. The gate-dependent spin-relaxation time further confirms the existence of the D'yakonov-Perel (DP) spin-relaxation mechanism and enhanced spin-orbit coupling under positive gate bias. These findings demonstrate electrostatic control of WL in $SrSnO_3$ and provide insights into phase-coherent transport and spin-relaxation process in disorder-dominated ultrawide bandgap oxide semiconductors, offering guidance for optimizing material growth and device operation.

## I. Introduction

The growing demand for efficient power conversion in data centers, electric vehicles, and power grids calls for semiconductor materials capable of operating at high voltages and power levels. Ultrawide bandgap (UWB) semiconductors offer great potential for such applications because their large bandgaps allow them to withstand high electric fields. However, combining a wide bandgap with high-mobility carriers remains challenging. Practical power devices based on UWB semiconductors also require controlled doping, interface engineering, and effective gate control[1]. Alkaline-earth stannates are promising oxide semiconductors with wide bandgaps and high electron mobility due to a dispersive Sn 5s conduction band[2,3]. Among them, strontium stannate ($SrSnO_3$, SSO) is particularly attractive because it balances a large direct bandgap of approximately 4.1 eV with high electron mobility[4,5]. Hybrid molecular beam epitaxy (hMBE) has enabled the growth of high-quality La-doped SSO films, with a room-temperature Hall mobility of 70 $cm^2/(Vs)$ at an electron density of $10^{20}$ $cm^{-3}$ in 12 nm films[6]. Top-gated metal-oxide-semiconductor field-effect transistors with SSO thin-film channels have achieved field-effect mobilities above 65 $cm^2/(Vs)$[7] and a large on/off current ratio over $10^{11}$.[8]

Despite the progress in materials and devices, understanding disorder-dominated electron transport remains essential to realize further advances in SSO. Although its dispersive conduction band favors high mobility, transport in real films is governed by scattering from dopants, defects, and interfaces. Low-temperature magnetotransport measurements provide a sensitive means of identifying scattering mechanisms and extracting characteristic length scales for scattering and phase coherence. By clarifying disorder-limited conduction, these studies can guide growth and doping strategies to reduce scattering and improve material quality, and inform device design in the diffusive transport regime[9].

Weak localization (WL) provides a framework for probing phase-coherent transport in disordered conductors. An electron can traverse closed scattering paths in opposite directions. Constructive interference between these time-reversed paths enhances the probability of returning to the starting point and thus localizes electrons, reducing the conductivity. An applied magnetic field introduces a relative phase shift between the two paths, suppressing the constructive interference and typically producing negative magnetoresistance at low fields[10,11]. Analysis of the magnetoconductance can therefore quantify the phase-coherence length as a function of temperature, providing insight into dephasing and the effective dimensionality of transport. For $SrSnO_3$ films, while WL has been reported in previous magnetotransport experiments[12-14], a comprehensive angle- and gate-dependent study is yet missing, and a deeper understanding of the electrostatic control over phase-coherent transport and the spin-relaxation process is urgently needed.

Here, we conducted magnetotransport measurements to investigate temperature, magnetic-field orientation, and gate control of the WL behavior in top-gated hMBE-grown La-doped SSO devices. Hall measurements before and after $Al_2O_3$ dielectric capping layer deposition reveal enhanced carrier density and mobility, suggesting a dielectric-induced doping effect that suppresses ionized impurity scattering. Temperature-dependent magnetoconductance is analyzed using the Iordanskii-Lyanda-Geller-Pikus (ILP) and Hikami-Larkin-Nagaoka (HLN) models[15-17], and the best fit using the ILP model yields a phase-coherence length of 112.1 nm at 1.7 K. Angle-dependent measurements show low-field collapse under perpendicular field scaling, supporting predominantly in-plane coherent transport. The magnetoconductance, phase-coherence length, and spin-relaxation length can further be modulated by a top gate, indicating the existence of the DP spin-relaxation mechanism and electron-electron interaction in La-doped SSO films at 1.7 K. This work provides further insight into the scattering and dephasing processes governing transport in $SrSnO_3$ thin films.

## II. Experimental methods

The La-doped SSO film with a thickness of 15.4 nm was deposited on a $GdScO_3$ substrate by radical-based hMBE. Sr and La were supplied from conventional effusion cells, with cell temperatures of 410 and 1100 °C, respectively. The Sr beam-equivalent pressure (BEP) was 1.86 × $10^{-8}$ mbar. Hexamethylditin (HMDT) was introduced through a vapor inlet as the Sn source; the HMDT Baratron pressure was 180 mTorr, and its BEP was 1.18 × $10^{-5}$ mbar. Activated oxygen was supplied by a 250 W radio-frequency plasma at an oxygen background pressure of 5 × $10^{-6}$ Torr. The high-resolution $2\theta$-$\omega$ X-ray diffraction scan and the rocking curve of the film $(002)_{pc}$ reflection are shown in Figs. 1(a) and 1(b), respectively.

After deposition, the La-doped SSO film was patterned into six-terminal Hall bars with channel dimensions of 125 µm × 25 µm. Electrical contacts (Cr (50 nm)/Au (50 nm)) were patterned using photolithography and subsequently deposited by electron-beam evaporation. A 20 nm $Al_2O_3$ dielectric was deposited by atomic layer deposition (ALD) at 200 °C, followed by Cr (50 nm) and Au (50 nm) top-gate patterning and deposition. Figs. 1(c) and 1(d) show the optical image of the top-gated Hall-bar device and the schematic of the cross-section of the channel.

Electrical transport measurements were performed on Hall-bar channels using a Quantum Design DynaCool cryostat system with a 9 T magnet and a rotator probe. An AC excitation of 100 nA was applied using a Keithley 6221 current source at 5.777 Hz. The AC voltage was measured using SR860 lock-in amplifiers synchronized to the AC current source. The DC gate voltage bias was applied using a Keithley 2450 voltage source meter. The detailed data symmetrization procedures based on the raw longitudinal and transverse resistance data are provided in Supplementary Note 1. All data shown below have been symmetrized.

## III. Results and discussion

We first compare the Hall measurement results on the same sample before (denoted as "non-ALD" hereafter) and after (denoted as "ALD-capped" hereafter) ALD deposition of the 20 nm $Al_2O_3$ dielectric layer. Fig. 2(a) shows the temperature dependence of the zero-field longitudinal resistivity ($\rho_{xx}$) for both non-ALD and ALD-capped channels. While both channels exhibit insulating-like temperature dependence with clear increases in resistivity at lower temperatures, the ALD-capped channel shows a significantly lower resistivity throughout the measured temperature range (Fig. 2(a)). To further understand the origin of the reduced resistivity after ALD capping, Hall measurements were performed to relate the change in longitudinal resistivity to the carrier density (Fig. 2(b)) and mobility (Fig. 2(c)). Within the single-carrier approximation, the effective Hall sheet density $n_{2D}$ and Hall mobility $\mu_H$ are extracted from the antisymmetrized Hall slope $R_H = dR_{xy}/dB$ and $\rho_{xx}(0)$:

$$n_{2D} = \frac{1}{e|R_H|}, \quad \mu_H = \frac{1}{en_{2D}\rho_{xx}(0)}, \tag{1}$$

where $e$ is the elementary charge. The ALD-capped channel shows an $n_{2D}$ of $2.76 \times 10^{13}$ cm$^{-2}$ at room temperature, which rapidly drops to $2.16 \times 10^{13}$ cm$^{-2}$ at 1.7 K, suggesting the freeze-out of thermally populated carriers at low temperatures. The Hall mobility of the ALD-capped channel decreases from 47.80 cm$^2$/(Vs) at room temperature to 37.37 cm$^2$/(Vs) at 1.7 K. By fitting mobility with a power-law temperature dependence:

$$\mu_H \propto T^{\alpha}, \tag{2}$$

we can extract a temperature coefficient of $\alpha = 0.063$. Such a small $\alpha$ indicates that carrier transport is dominated by temperature-insensitive scattering mechanisms, in this case likely surface roughness scattering[18].

The non-ALD channel, on the other hand, exhibits a substantially lower $n_{2D}$ of $1.70 \times 10^{13}$ cm$^{-2}$ and lower Hall mobility of 28.66 cm$^2$/(Vs) at room temperature compared to the ALD-capped channel. This difference is consistent with previous reports of dielectric-induced modulation doping in ALD-capped SSO devices[19]. The fixed charges in the $Al_2O_3$ layer can induce additional carriers in the channel, and the higher density in the capped channel can strengthen the screening of charged impurities and thereby enhance mobility[9]. Upon cooling, the mobility of the non-ALD channel decreases with a fitted temperature dependence parameter of $\alpha = 0.504$, which can be attributed to a combination of surface roughness scattering ($\alpha \approx 0$) and ionized impurity scattering ($\alpha \approx 1.5$). Meanwhile, the extracted carrier density of the non-ALD channel increases at low temperature. However, the low-temperature data for this non-ALD channel should be interpreted with caution, as its high resistance makes the measurements susceptible to experimental artifacts. These possible artifacts include the channel resistance approaching the input resistance of the lock-in amplifier (10 MΩ), an increasing out-of-phase component in the AC measurements, and uncertainty in the symmetrization procedure arising from substantial mixing of the longitudinal resistance into the transverse Hall resistance signal. Therefore, we restrict the subsequent WL analysis to the ALD-capped channel.

Fig. 3(a) shows the magnetic field dependence of the magnetoconductance $\Delta\sigma_{xx}$ from 300 K to 1.7 K. A large positive magnetoconductance was observed at low temperature and can be attributed to WL behavior. The WL feature becomes larger and sharper near zero field as temperature decreases, consistent with a growing phase-coherent contribution[10,11]. The WL-induced positive magnetoconductance gradually diminishes with increasing temperature due to decoherence as scattering increases.

To quantitatively determine the phase-coherence length, the WL magnetoconductance curves measured at 1.7, 5, 10, and 20 K were fitted using the linear Iordanskii-Lyanda-Geller-Pikus (ILP) model with a spin-orbit coupling (SOC) term[20], as shown in Fig. 3(b). The fits were restricted to a magnetic field range of 4 T to limit contributions from ordinary orbital magnetoresistance, which may become significant at higher fields. In addition, fits using the Hikami-Larkin-Nagaoka (HLN) model are performed and presented in Supplementary Note 2[17]. While both models describe the WL behavior reasonably well within the fitting range, the ILP model provides a slightly more accurate overall fit to the experimental data. This suggests that ILP better describes D'yakonov-Perel (DP) momentum-dependent spin relaxation[21,22], as Rashba-type SOC is expected in the system due to structural asymmetry. We therefore use the ILP model for the subsequent analysis.

The ILP fitting of field-dependent magnetoconductance can be expressed as:

$$\begin{aligned} &\Delta\sigma^{a}_{\mathrm{xx}}(B) = \sigma_{\mathrm{xx}}(B) - \sigma_{\mathrm{xx}}(B_a), \\ &\Delta\sigma^{a}_{\mathrm{ILP}}(B) = A\frac{e^2}{2\pi^2\hbar}\left[F_{\mathrm{ILP}}\left(\frac{B_\phi}{|B|},\frac{B_{\mathrm{so}}}{|B|}\right) - F_{\mathrm{ILP}}\left(\frac{B_\phi}{|B_a|},\frac{B_{\mathrm{so}}}{|B_a|}\right)\right], \end{aligned} \tag{3}$$

where $A_{\mathrm{ILP}}$ is empirical, $\hbar$ is the reduced Planck constant, $F_{\mathrm{ILP}}$ is the zero-field-subtracted ILP function evaluated using the linear-SOC limit[20], $B_\phi$ is the phase-coherence field, and $B_{\mathrm{so}}$ is the characteristic spin-orbit field. Both the experimental data and model are referenced to $B_a = 0.1$ T.

Writing $r = B_\phi / |B|$ and $s = B_{\mathrm{so}} / |B|$, a closed form for the field-dependent term is

$$F_{\mathrm{ILP}}(B) = \frac{1}{2}\left[\mathrm{Re}\sum_{j=1}^{3} u_j\psi\left(\frac{1}{2}+r-v_j\right) - \psi\left(\frac{1}{2}+r\right) - 2\ln r + \frac{4}{4(r+s)^2-1}\right], \tag{4}$$

where $\psi$ is the digamma function[20]. The three values of $v_j$ describe the different modes in the ILP model, and $u_j$ determines the contribution of each mode. Their mathematical definitions and details of the numerical calculation are given in Supplementary Note 2. The fitted phase-coherence field $B_\phi$ (characteristic spin-orbit field $B_{\mathrm{so}}$) can be converted to phase-coherence length $L_\phi$ (spin-relaxation length $L_{\mathrm{so}}$) and phase-relaxation time $\tau_\phi$ (spin-relaxation time $\tau_{\mathrm{so}}$):

$$\begin{aligned} &\tau_x = \frac{\hbar}{4eDB_x}, \\ &L_x = \sqrt{D\tau_x} = \sqrt{\frac{\hbar}{4eB_x}},\ x = \phi,\ \mathrm{so}, \end{aligned} \tag{5}$$

where $D$ is the diffusion constant and the detailed description is shown in Supplementary Note 3.

The ILP-derived $L_\phi$ decreases from 112.1 ± 4.8 nm at 1.7 K to 53.78 ± 0.47, 34.85 ± 0.16, and 23.27 ± 0.14 nm at 5, 10, and 20 K, respectively (Fig. 3(c)). The temperature dependence of the spin-relaxation length $L_{so}$ is shown in Fig. S4. The $L_{so}$ decreases from 39.6 to 28.2 nm as the temperature increases from 1.7 to 20 K. The $L_\phi$ follows a power law of temperature dependence:

$$L_\phi(T) \propto T^{-p/2}, \tag{6}$$

which yields a temperature exponent $p$ = 1.32 ± 0.03. This $p$ falls between the expected values for electron-electron dephasing in the two-dimensional (2D) ($p$ = 1) and three-dimensional (3D) ($p$ = 1.5) diffusive limits, which is reasonable for a film with finite film thickness (15.4 nm) comparable to $L_\phi$.

We further measured the WL response as a function of magnetic field orientation. Fig. 4(a) shows the magnetoconductance curves measured at 1.7 K with the field tilted by 15°, 30°, 45°, 60°, and 75° relative to the film normal. The magnetoconductance at the maximum applied field, $\Delta\sigma_{xx}$(9 T), decreases as the field is tilted toward the film plane. To analyze this angle dependence, the magnetic field is decomposed into in-plane and out-of-plane components[10,17]:

$$\begin{aligned} B_\perp &= \left|B\cos\theta\right|, \\ B_\parallel &= \left|B\sin\theta\right|. \end{aligned} \tag{7}$$

Fig. 4(b) replots the magnetoconductance curves at each tilt angle as a function of the perpendicular field $B_\perp$. All the curves collapse under low $B_\perp$, indicating that the WL response is governed primarily by the perpendicular field. This behavior is expected because the dominant interference paths lie within the film plane and are therefore primarily sensitive to the perpendicular magnetic field component. At higher fields, however, the curves measured at larger tilt angles saturate at lower magnetoconductance values. At fixed $B_\perp$, the in-plane field component increases with tilt angle, suggesting that the high-field deviations reflect an additional in-plane-field-dependent contribution. Possible origins include in-plane-field-induced magnetoresistance and finite-thickness orbital effects. In the latter case, interference paths extending in the film thickness direction can enclose magnetic flux from the in-plane field, modifying the WL response.

To elucidate the effects of carrier density on WL, gate-dependent magnetoconductance was further investigated (on a separate device). The magnetoconductance curves were measured under the gate voltages $V_g$ = 0, 0.5, 1, 2, 3, and 4 V at 1.7 K, and the ILP fits were performed at every gate bias with the magnetic field up to 4 T, as shown in Fig. 5(a). Using Eq. (5), the extracted $L_\phi$ ($L_{so}$) increases from 48.05 ± 1.05 nm (38.54 ± 0.34 nm) at $V_g$ = 0 V to 67.54 ± 1.34 nm (44.46 ± 0.37 nm) at $V_g$ = 4 V, as shown in Fig. 5(b). The extracted phase-coherence time $\tau_\phi$ shows a weak gate dependence between 0 and 4 V, with an average value of 12.6 ps (Fig. 5(c)). According to Eq. (5), $L_\phi$ depends on the carrier diffusion coefficient $D$ and the phase-coherence time $\tau_\phi$. The nearly gate-independent $\tau_\phi$ indicates that the increase in $L_\phi$ under positive gate bias is due to enhanced carrier diffusion $D$ as the channel conductivity increases (as shown in Fig. S3), implying that electron-electron scattering is the main phase-relaxation mechanism. Similar gate-dependent trends have been reported in other 2D semiconductors such as $MoS_2$[23] and *n*-type tellurium[24], where gate-dependent coherence was interpreted as electron-electron-interaction-induced dephasing. According to the Altshuler-Aronov-Khmelnitsky (AAK) theory[25], the phase-coherence length $L_\phi$ associated with inelastic electron-electron scattering with small energy transfer can be expressed as[26]

$$L_\phi = \frac{h^2\sigma_{xx}}{\pi e^2}\left(m^* k_B T \cdot \ln\frac{\sigma_{xx} h}{e^2}\right)^{-1/2}, \tag{8}$$

where $m^*$ is the electron effective mass, and $k_B$ is the Boltzmann constant. Fig. 5(d) shows the $L_\phi$ as a function of the zero-field sheet conductance $\sigma_{xx}$ at 1.7 K. The linear dependence between $L_\phi$ and $\sigma_{xx}$ fitted by the AAK theory indicates that the electron-electron interaction is the dominant phase-relaxation mechanism in La-doped SSO films under different gate voltages at 1.7 K[24].

The ILP model further provides insight into the gate-tunable SOC effect and spin relaxation process in the WL effect. Over the same gate-voltage range, the spin-relaxation length $L_{so}$ increases moderately from 38.54 nm to 44.46 nm as $V_g$ increases from 0 to 4 V (Fig. 5(b)), meanwhile the spin-relaxation time $\tau_{so}$ decreases from 5.7 to 4.1 ps (Fig. 5(c)). This discrepancy can be explained by the nearly twofold increase in $D$, which allows carriers to diffuse further within the shorter $\tau_{so}$ according to Eq. (5). A decrease in $\tau_{so}$ also suggests a stronger spin-orbit interaction at a higher gate voltage, which is expected as Rashba-type spin-orbit interaction will be enhanced at higher electric fields. For the DP spin relaxation driven by linear Rashba coupling, the spin-relaxation rate scales as[20]

$$\frac{1}{\tau_{so}} \propto \left(\frac{\alpha_R k_F}{\hbar}\right)^2 \tau_{tr}, \tag{9}$$

where $\alpha_R$ is the Rashba coefficient, $k_F$ is the Fermi wave vector, and $\tau_{tr}$ is the momentum-scattering time (detailed in Supplementary Notes 3 and 5). The relation between spin-relaxation time $\tau_{so}$ and momentum-scattering rate $1/\tau_{tr}$ is shown in Fig. S5. The linear fitting indicates that $\tau_{so}$ is inversely proportional to $\tau_{tr}$, implying the existence of the DP spin-relaxation mechanism in UWB SSO films[24], further justifying the use of ILP over HLN for WL fitting, especially under strong gate bias.

## IV. Conclusions

In summary, we have comprehensively investigated the effects of dielectric capping, temperature, magnetic-field orientation, and gate bias on electron transport in top-gated La-doped $SrSnO_3$ thin-film devices. A large positive magnetoconductance was observed at low temperatures, characteristic of WL. An ILP analysis yields effective phase-coherence lengths ranging from 23.3 to 112.1 nm over 20-1.7 K. The smaller fitting residuals of the ILP model compared to the HLN model suggest the presence of SOC, likely of the Rashba type, due to the asymmetric film stack. The WL curves under different magnetic field orientations collapse when plotted against the perpendicular field component, pointing to quasi-2D transport in the thin-film limit. Furthermore, a top gate enables modulation of channel conduction and electron mobility, which further affects the WL and phase-coherence length, consistent with enhanced inelastic electron-electron scattering that leads to dephasing as the channel conductivity increases. The gate dependence of WL further reveals electrostatic control of phase-coherent transport and the spin relaxation process. The results support electron-electron-interaction-driven dephasing and D'yakonov-Perel' spin relaxation, with gate-enhanced spin-orbit coupling. These results establish quantitative gate modulation of disorder-dominated transport in a UWB oxide channel, offering guidance for optimizing material growth, interface engineering, and device operation.

## Figures

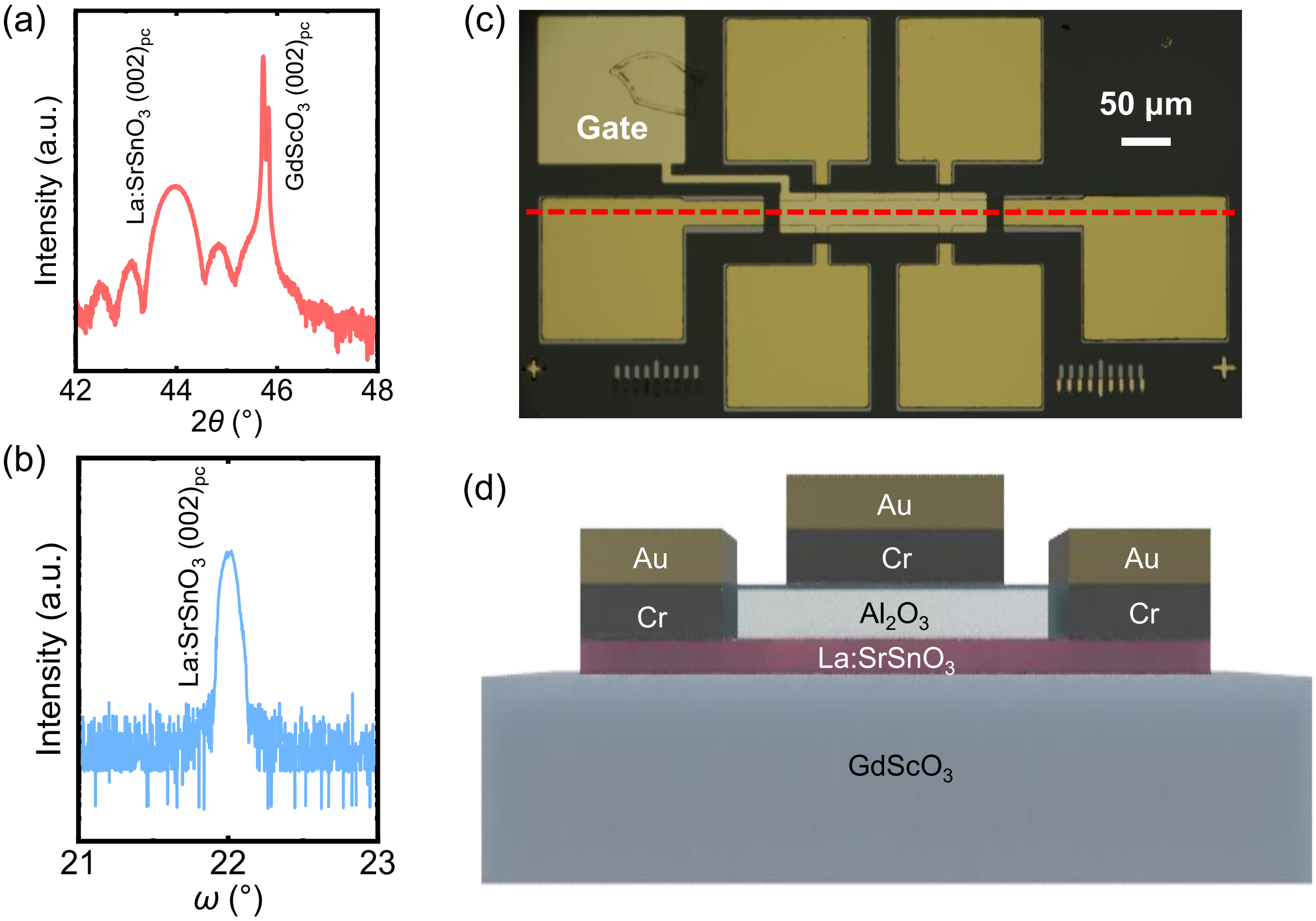

**FIG. 1.** Structural characterization and device layout of the La-doped $SrSnO_3$ film. (a) High-resolution $2\theta$-$\omega$ X-ray diffraction scan around the pseudocubic (002) reflections of the film and $GdScO_3$ substrate. (b) Rocking curve of the La-doped $SrSnO_3$ $(002)_{pc}$ reflection. (c) Optical image of the fabricated top-gated Hall-bar device. (d) Cross-sectional schematic along the red dashed line in (c).

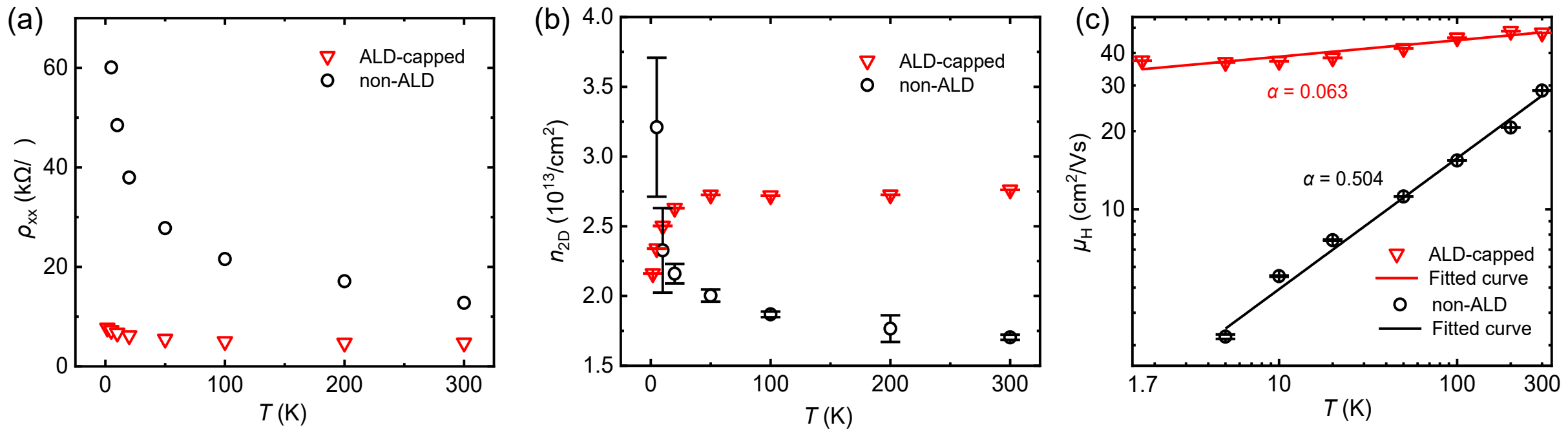

**FIG. 2.** Temperature-dependent transport measurements on ALD-capped and non-ALD samples. (a) Temperature dependence of the zero-field longitudinal resistivity $\rho_{xx}$ for the ALD-capped (red open inverted triangles) and non-ALD (black open circles) samples. (b) Temperature dependence of the Hall sheet density $n_{2D}$ for the ALD-capped and non-ALD samples. Error bars stem from the standard errors of the Hall slope fits. (c) Temperature dependence of the Hall mobility (double logarithmic scales) with the power-law fits of Eq. (2): $\alpha$ = 0.063 and 0.504 for ALD-capped and non-ALD samples, respectively. Error bars are from the standard errors of the Hall slope fits.

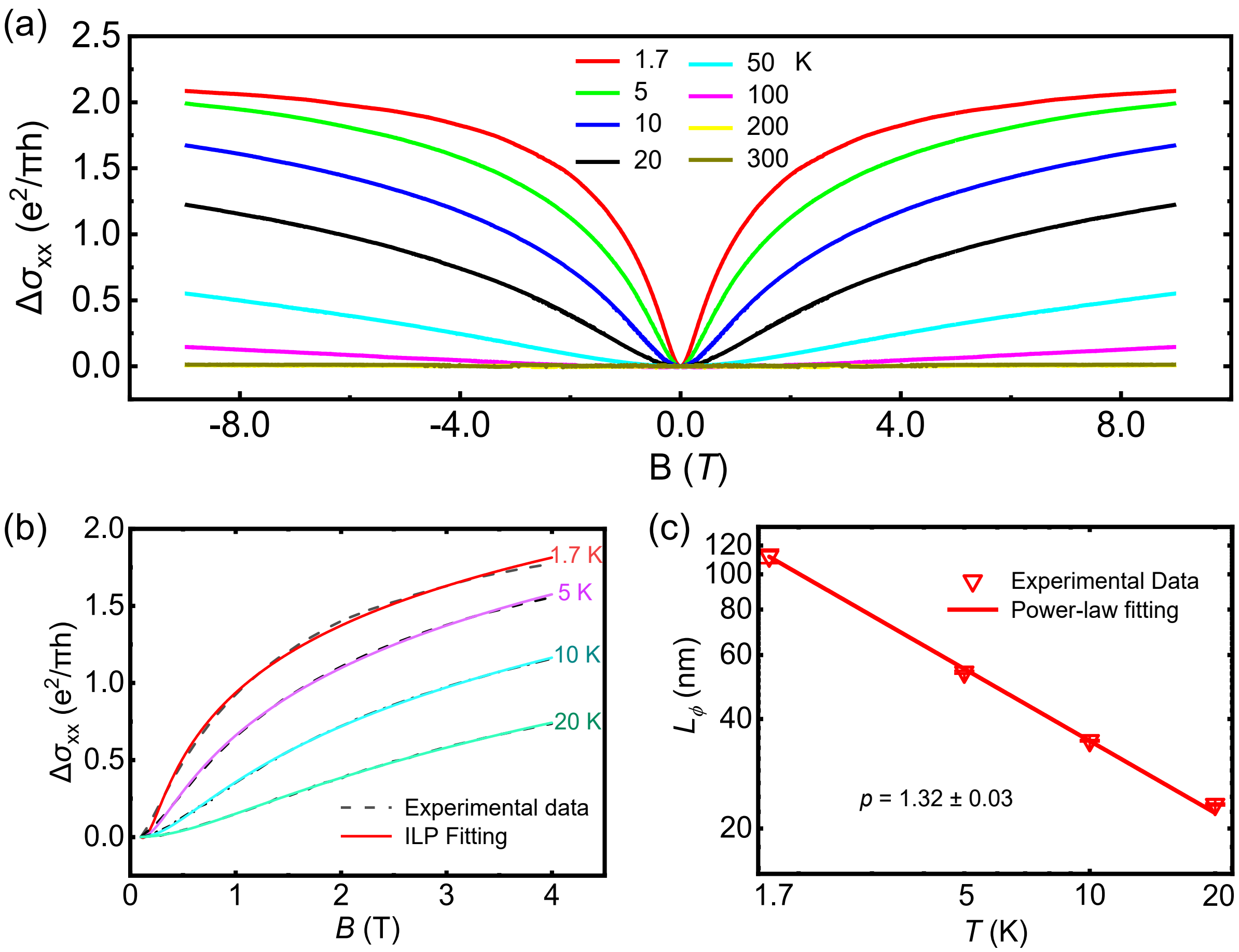

**FIG. 3.** Magnetic-field-dependent magnetoconductance and ILP analysis at different temperatures. (a) Magnetic field dependence of magnetoconductance at 1.7, 5, 10, 20, 50, 100, 200, and 300 K, respectively. (b) Experimental curves (dashed) and linear ILP fits (solid) at 1.7, 5, 10, and 20 K. The fitting range is from 0.11 to 4 T. (c) ILP-derived $L_\phi$ (red inverted triangles) and the temperature power-law fit of Eq. (6), with $p$ = 1.32 ± 0.03. Error bars are from the standard errors of the ILP fits.

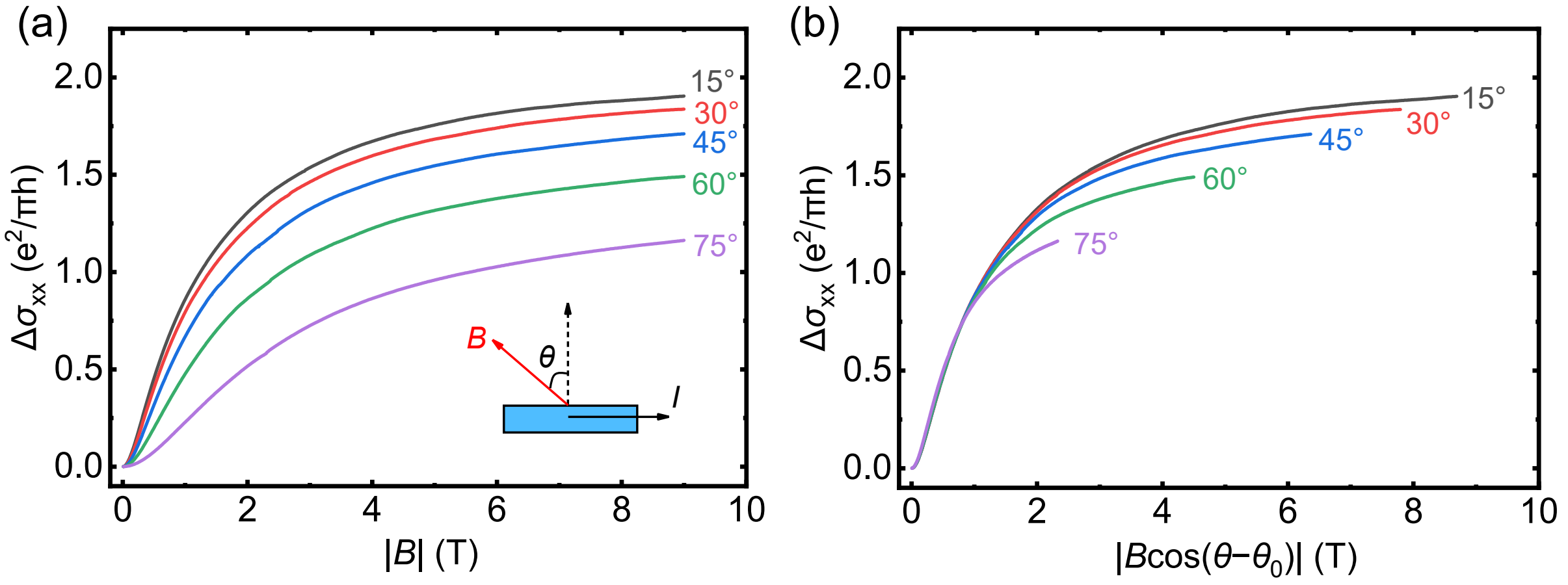

**FIG. 4.** Magnetic field angle dependence of the magnetoconductance at 1.7 K. (a) Magnetic field magnitude dependence of the magnetoconductance under different field orientations of 15°, 30°, 45°, 60°, and 75°. The inset defines the magnetic field angle $\theta$ relative to the film normal and the in-plane current direction $I$. (b) Magnetoconductance versus $|B\cos\theta|$. Each curve stops at its own measured projected endpoint.

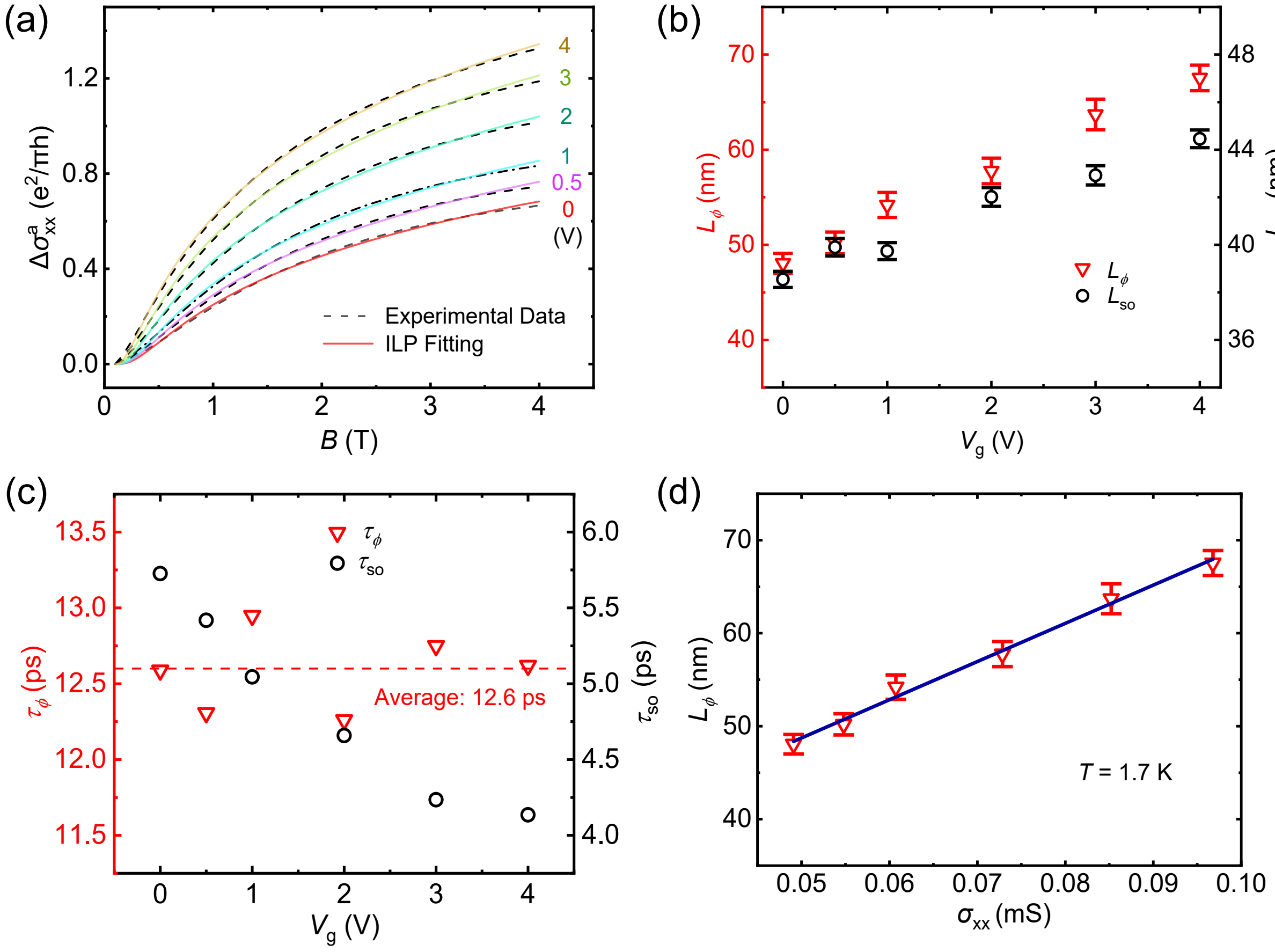

**FIG. 5.** Gate-dependent magnetotransport at 1.7 K under zero magnetic field. (a) Experimental magnetoconductance (dashed) and ILP fits (solid) under gate voltage $V_g$ = 0, 0.5, 1, 2, 3, and 4 V with the fitting range from 0.12 to 4 T. (b) Gate dependence of the $L_\phi$ (left) and $L_{so}$ (right) from ILP fitting. Error bars are from the standard errors of the ILP fits. (c) Gate dependence of the $\tau_\phi$ (left) and $\tau_{so}$ (right). (d) $L_\phi$ as a function of sheet conductance $\sigma_{xx}$ at 1.7 K. The red line is the linear fitting result according to the AAK theory.

**Supplementary material**

The supplementary material contains the resistance symmetrization and conductivity calculation, the ILP and HLN models, detailed description of the diffusion constant, temperature dependence of spin-relaxation length, and the relation between spin-relaxation time and momentum-scattering rate.

**Acknowledgements**

This research was supported by the Nano & Material Technology Development Program through the National Research Foundation of Korea (NRF) funded by the Ministry of Science and ICT (RS-2024-00460372). G. Qiu acknowledges the Department of Electrical and Computer Engineering and the College of Science and Engineering at the University of Minnesota, Twin Cities for the start-up funding support. Parts of this work were carried out at the Characterization Facility, University of Minnesota, which receives partial support from the NSF through UMNMRSEC. Portions of this work were carried out at the Minnesota Nano Center, which receives support from the NSF through the National Nanotechnology Coordinated Infrastructure (NNCI) under Award No. ECCS-2025124.

**Author Declarations**

**Conflict of Interest:** The authors have no conflicts to disclose.

**Author Contributions:**

**Long Cheng:** Conceptualization (lead); Data curation (lead); Formal analysis (lead); Investigation (lead); Writing – original draft (lead); Writing – review & editing (lead).

**Junghyun Koo:** Conceptualization (equal); Data curation (equal); Formal analysis (equal); Investigation (equal); Writing – original draft (equal); Writing – review & editing (equal).

**Dongho Won:** Conceptualization (equal); Data curation (equal); Formal analysis (equal); Investigation (equal); Writing – original draft (equal); Writing – review & editing (equal).

**Donghwan Kim:** Conceptualization (equal); Data curation (equal); Formal analysis (equal); Investigation (equal); Writing – original draft (equal); Writing – review & editing (equal).

**Chengyu Zhu:** Conceptualization (equal); Data curation (equal); Formal analysis (equal); Investigation (equal); Writing – original draft (equal); Writing – review & editing (equal).

**Bharat Jalan:** Conceptualization (equal); Data curation (equal); Formal analysis (equal); Investigation (equal); Writing – original draft (equal); Writing – review & editing (equal).

**Kiyoung Lee:** Conceptualization (equal); Data curation (equal); Formal analysis (equal); Investigation (equal); Writing – original draft (equal); Writing – review & editing (equal).

**Weideng Sun:** Conceptualization (equal); Data curation (equal); Formal analysis (equal); Investigation (equal); Writing – original draft (equal); Writing – review & editing (equal).

**Gang Qiu:** Conceptualization (equal); Funding acquisition (lead); Project administration (lead); Data curation (equal); Formal analysis (equal); Investigation (equal); Supervision (lead); Writing – review & editing (equal).

## Data Availability

The data that support the findings of this study are available from the corresponding author upon reasonable request.